\documentclass[11pt, letter]{article}
\usepackage[top=1in,bottom=1in,left=1in,right=1in]{geometry}
\usepackage[T1]{fontenc}
\usepackage{slashed}
\usepackage{amsmath,amssymb,amsfonts}
\usepackage{epsfig}
\usepackage{comment}
\usepackage{cancel}
\usepackage{bbm}
\usepackage{array}
\usepackage{color}
\usepackage{framed}
\usepackage{graphicx}
\usepackage{mathrsfs}
\usepackage{titlesec}
\usepackage[dotinlabels]{titletoc}
\usepackage{wrapfig}
\usepackage[all]{xy}
\usepackage{young}
\usepackage[vcentermath]{youngtab}
\usepackage{hyperref}
\hypersetup{colorlinks=true}
\hypersetup{linkcolor=blue}
\hypersetup{citecolor=blue}
\hypersetup{urlcolor=blue}
\numberwithin{equation}{section}
\usepackage[utf8]{inputenc}
\usepackage{slashed}
\usepackage{cite}
\usepackage{mathtools}

\def\e{{\epsilon}}
\def\ve{{\varepsilon}}

\def\CA{{\mathcal A}}

\def\CC{{\mathcal C}}

\def\CI{{\mathcal I}}
\def\ci{{\mathcal I}}
\def\CJ{{\mathcal J}}

\def\CL{{\mathcal L}}
\def\CM{{\mathcal M}}

\def\CO{{\mathcal O}}
\def\CP{{\mathcal P}}

\def\CS{{\mathcal S}}

\def\SC{{\mathscr C}}

\def\p{\partial}
\newcommand{\avg}[1]{\langle \,#1\, \rangle}

\def\mdd{{\mathbb D}}

\def\mrr{{\mathbb R}}

\def\mzz{{\mathbb Z}}
   \makeatletter
  \let\over=\@@over \let\overwithdelims=\@@overwithdelims
  \let\atop=\@@atop \let\atopwithdelims=\@@atopwithdelims
  \let\above=\@@above \let\abovewithdelims=\@@abovewithdelims
\renewcommand\section{\@startsection {section}{1}{\z@}%
                                   {-3.5ex \@plus -1ex \@minus -.2ex}%nn
                                   {2.3ex \@plus.2ex}%
                                   {\normalfont\large\bfseries}}
\renewcommand\subsection{\@startsection{subsection}{2}{\z@}%
                                     {-3.25ex\@plus -1ex \@minus -.2ex}%
                                     {1.5ex \@plus .2ex}%
                                     {\normalfont\bfseries}}
\def\wt{\widetilde}

\def\0{{(0)}}
\def\1{{(1)}}
\def\2{{(2)}}
\def\3{{(3)}}
\def\4{{(4)}}
\def\+{{(+)}}
\def\-{{(-)}}

\def\ads{{\text{AdS}}}

\def\e{{\epsilon}}

\def\p{\partial}

\def\0{{(0)}}
\def\1{{(1)}}
\def\2{{(2)}}

\def\<{\langle }
\def\>{\rangle }

\def\bea{\begin{eqnarray}}
\def\eea{\end{eqnarray}}
\def\be{\begin{equation}}
\def\ee{\end{equation}}
\def\ba{\begin{align}}
\def\ea{\end{align}}
\def\beq{\begin{equation}}
\def\eeq{\end{equation}}
\def\beqa{\begin{eqnarray}}
\def\eeqa{\end{eqnarray}}
\def\beqar{\begin{eqnarray*}}

\def\[{\big[}
\def\]{\big]}

\def\p{\partial}

\begin{document}
\begin{titlepage}
\unitlength = 1mm
\ \\
\vskip 3cm
\begin{center}
%%%%%%%%%%%%%%%%%%%%---------------------TITLE---------------------%%%%%%%%%%%%%%%%%%%%%%%%

{\LARGE{\textsc{A $d$-Dimensional Stress Tensor for Mink$_{d+2}$ Gravity}}}

\vspace{0.8cm}
%%%%%%%%%%%%%%%%%%%%---------------------AUTHOR(S)---------------------%%%%%%%%%%%%%%%%%%%%%
Daniel Kapec$^\dagger$ and Prahar Mitra$^{\dagger *}$

\vspace{1cm}

{\it  $^\dagger$Center for the Fundamental Laws of Nature, Harvard University,\\
Cambridge, MA 02138, USA}

{\it  $^*$School of Natural Sciences, Institute for Advanced Study,\\
 Princeton, NJ 08540, USA}

\vspace{0.8cm}

%%%%%%%%%%%%%%%%%%%%---------------------ABSTRACT---------------------%%%%%%%%%%%%%%%%%%%%%%
\begin{abstract}
We consider the tree-level scattering of massless particles in $(d+2)$-dimensional asymptotically flat spacetimes. The $\CS$-matrix elements are recast as correlation functions of local operators living on a space-like cut $\CM_d$ of the null momentum cone. The Lorentz group $SO(d+1,1)$ is nonlinearly realized as the Euclidean conformal group on $\CM_d$. Operators of non-trivial spin arise from massless particles transforming in non-trivial representations of the little group $SO(d)$, and distinguished operators arise from the soft-insertions of gauge bosons and gravitons. The leading soft-photon operator is the shadow transform of a conserved spin-one primary operator $J_a$, and the subleading soft-graviton operator is the shadow transform of a conserved spin-two symmetric traceless primary operator $T_{ab}$. The universal form of the soft-limits ensures that $J_a$ and $T_{ab}$ obey the Ward identities expected of a conserved current and energy momentum tensor in a Euclidean CFT$_d$, respectively.
 \end{abstract}

\vspace{1.0cm}
\end{center}
\end{titlepage}
\pagestyle{empty}
\pagestyle{plain}
\pagenumbering{arabic}
 %%%%%%%%%%%%%%%%---------------------END OF TITLE PAGE AND ABSTRACT---------------------%%%%%%%%%%%%%%%%%%%%%%

\tableofcontents

\section{Introduction  }
The light-like boundary of asymptotically flat spacetimes, $\CI = \CI^+ \cup \CI^-$,  is a null cone with a (possibly singular) vertex at spatial infinity. Massless excitations propagating in such a spacetime pass through $\ci$ at isolated points on the celestial sphere. Guided by the holographic principle, one might hope that the $\CS$-matrix for the scattering of massless particles in asymptotically flat spacetimes in $(d+2)$-dimensions might be reexpressed as a collection of correlation functions of local operators on the celestial sphere $S^{d}$ at null infinity, with operator insertions at the points where the particles enter or exit the spacetime. The Lorentz group would then be realized as the group of conformal motions of the celestial sphere, and the Lorentz covariance of the $\CS$-matrix would guarantee that the local operators have well-defined transformation laws under the action of the Euclidean conformal group $SO(d+1,1)$.  On these general grounds one expects the massless $\CS$-matrix to display some of the features of a  $d$-dimensional Euclidean conformal field theory (CFT$_d$). 

It has recently become possible to make some of these statements more precise in four dimensions, due in large part to Strominger's infrared triangle that relates soft theorems, asymptotic symmetry groups and memory effects \cite{Barnich:2010eb,Strominger:2013lka,Strominger:2013jfa,He:2014laa,Adamo:2014yya,Geyer:2014lca,Kapec:2014opa,He:2014cra,Lysov:2014csa,Campiglia:2014yka,Strominger:2014pwa,Kapec:2014zla,Mohd:2014oja,Campiglia:2015yka,Kapec:2015vwa,He:2015zea,Campiglia:2015qka,Kapec:2015ena,Strominger:2015bla,Campiglia:2015kxa,Dumitrescu:2015fej,Kapec:2016aqd,Campiglia:2016jdj,Campiglia:2016hvg,Conde:2016csj,Gabai:2016kuf,Campiglia:2016efb,Kapec:2016jld,Conde:2016rom,Pasterski:2016qvg,He:2017fsb,Strominger:2017zoo,Campiglia:2017dpg,Nande:2017dba,Pasterski:2017kqt,Kapec:2017tkm,Pasterski:2017ylz,Mao:2017wvx,Choi:2017bna,Laddha:2017vfh}. While the specific details of a putative holographic formulation are expected to be model dependent,  it should be possible to make robust statements (primarily regarding symmetries) based on universal properties of the $\CS$-matrix. One interesting  class of universal statements about the $\CS$-matrix concerns the so-called soft-limits \cite{Bloch:1937pw,Low:1954kd,GellMann:1954kc,Low:1958sn,Weinberg:1965nx,Burnett:1967km,Gross:1968in,Jackiw:1968zza} of scattering amplitudes. In the limit when the wavelength of an external gauge boson or graviton becomes much larger than any scale in the scattering process, the $\CS$-matrix factorizes into a universal soft operator (controlled by the soft particle and the quantum numbers of the hard particles) acting on the amplitude without the soft insertion. This sort of factorization is reminiscent of a Ward identity, and indeed in four dimensions the soft-photon, soft-gluon, and  soft-graviton theorems have been recast in the form of Ward identities for conserved operators in a putative CFT$_2$ \cite{Nair:1988bq,Strominger:2013lka,He:2015zea,Cheung:2016iub,Kapec:2016jld,He:2017fsb,Nande:2017dba}. Most importantly for the present work, in \cite{Kapec:2016jld,He:2017fsb, Cheung:2016iub} an operator was constructed from the subleading  soft-graviton theorem whose insertion into the four dimensional $\CS$-matrix reproduces the Virasoro Ward identities of a CFT$_2$ energy momentum tensor. The subleading  soft-graviton theorem holds in all dimensions \cite{Cachazo:2014fwa,Schwab:2014xua,Bern:2014oka,Broedel:2014fsa,Sen:2017xjn,Sen:2017nim,Chakrabarti:2017ltl}, so it should be possible to construct an analogous operator in any dimension. We will see that this is indeed the case, and that the construction is essentially fixed by Lorentz (conformal) invariance.

The organization of this paper is as follows. In section \ref{sec:kinematics} we establish our conventions for massless particle kinematics and describe the map from the $(d+2)$-dimensional $\CS$-matrix to a set of $d$-dimensional ``celestial correlators'' defined on a space-like cut of the null momentum cone. Section \ref{sec:lorentztransform} describes the realization of the Euclidean conformal group on these correlation functions in terms of the embedding space formalism. Section \ref{sec:currents} outlines the construction of conserved currents -- namely the conserved $U(1)$ current and the stress tensor -- in the boundary theory and their relations to the leading soft-photon and subleading soft-graviton theorems. Section \ref{sec:comments} concludes with a series of open questions.  In appendix \ref{app:coordinates}, we briefly discuss the bulk space-time interpretation of our results and their relations to previous work. 

\section{Massless Particle Kinematics }\label{sec:kinematics}
The basic observable in asymptotically flat quantum gravity is the $\CS$-matrix element 
\be
\CS = \langle \;  \text{out} \; |\;   \text{in} \; \rangle
\ee
 between an incoming state on past null infinity $(\CI^-)$ and an outgoing state on future null infinity $(\CI^+)$. The perturbative scattering states in asymptotically flat spacetimes are characterized by  collections of well separated, non-interacting particles.\footnote{In four dimensions, the probability to scatter into a state with a finite number of gauge bosons or gravitons is zero due to infrared divergences \cite{Bloch:1937pw}. In higher dimensions, infrared divergences are absent and one can safely consider the usual Fock space basis of scattering states.} Each massless particle is characterized by a null momentum $p^\mu$ and a representation of the little group $SO(d)$, as well as a collection of other quantum numbers such as charge, flavor, etc.
Null momenta are constrained to lie on the future light cone $\CC^+$ of the origin in momentum space $\mrr^{d+1,1}$,
\be\label{cpdef}
\CC^+= \{ p^\mu \in \mathbb{R}^{d+1,1} \,\big|\, p^2=0 \, , p^0 > 0  \} \; .
\ee
 A convenient parametrization for the momentum, familiar from the embedding space formalism in conformal field theories \cite{SimmonsDuffin:2012uy,Penedones:2016voo}, is given by 
\be \label{momentum}
p^\mu(\omega,x)=\omega \Omega(x)\hat{p}^\mu(x) \; , \hspace{.5 in} \hat{p}^\mu(x)=  \left(\frac{1+x^2}{2},x^a,\frac{1-x^2}{2} \right) \; , \hspace{.25 in} \omega \geq 0\;, \;\; x^a \in \mathbb{R}^d \; ,
\ee
where $x^2= x^a x_a  = \delta_{ab}x^a x^b  $. The metric on this null cone is degenerate and is given by
\be
ds^2_{\CC^+} =dp^\mu dp_\mu= 0 d\omega^2 + \omega^2\Omega(x)^2 dx^a dx_a \; .
\ee

For a fixed $\omega$, this parametrization specifies a $d$-dimensional space-like cut $\CM_d$ of the future light cone with a conformally flat metric induced from the flat Lorentzian metric on $\mathbb{R}^{d+1,1}$:
\be\label{Mdmetric}
ds_{\CM_d}^2 = \Omega(x)^2 dx^a dx_a  \;. 
\ee
$\Omega(x)$ defines the conformal factor on $\CM_d$. In most of what follows we will choose $\Omega(x)=1$ for computational simplicity, although the generalization to an arbitrary conformally flat Euclidean cut is straightforward.\footnote{Other choices of $\Omega(x)$ have also proved useful in previous analyses. In particular, the authors of \cite{He:2014laa,Kapec:2015ena,Cachazo:2014fwa,He:2014cra,Strominger:2013jfa} choose $\Omega(x)=2(1+x^2)^{-1}$, yielding the round metric on $S^d$. 
For non-constant $\Omega(x)$, $d$-dimensional partial derivatives are simply promoted to covariant derivatives, and powers of the Laplacian are replaced by their conformally covariant counterparts, the GJMS operators \cite{JLMS:JLMS0557}. }
 The $(d+2)$-dimensional Lorentz-invariant measure takes the form
\be
\int \frac{d^{d+1}p}{p^0} = \int  d^dx\int d\omega \omega^{d-1}\; ,
\ee
while the Lorentzian inner product is given by
\be
- 2 \hat{p}(x_1)\cdot \hat{p}(x_2)=(x_1-x_2)^2 \; .
\ee
Massless particles of spin $s$ can be described by symmetric traceless fields
\be
\Phi_{\mu_1 \dots \mu_s}(X) = \sum_{a_i}\int \frac{d^{d+1}p}{(2\pi)^{d+1}}\frac{1}{2\omega_p}\ve_{\mu_1 \ldots \mu_s}^{a_1\ldots a_s}(p) \left[	\CO_{a_1\dots a_s}(p)e^{ip \cdot X} + \CO^\dagger_{a_1\dots a_s}(p) e^{-ip\cdot X} \right] 
\ee
satisfying the equations
\be
\Box_X\Phi_{\mu_1 \dots \mu_s}(X) = 0 \; , \hspace{.5 in} \p_{\mu_1 }\Phi^{\mu_1}{}_{\mu_2 \dots \mu_s}(X)=0 \; . 
\ee
Under gauge transformations,
\be
\Phi_{\mu_1 \dots \mu_s}(X) ~ \to~  \Phi_{\mu_1 \dots \mu_s}(X) + \p_{(\mu_1} \lambda_{\mu_2 \dots \mu_s)}(X) \; . 
\ee 
We will work in the gauge
\be
n^{\mu_{1}}\Phi_{\mu_1 \dots \mu_s}(X) = 0 \; , \hspace{.5 in} n^\mu = (1,0^a,-1) \; .
\ee
A natural basis for the vector representation of the little group $SO(d)$ is given in terms of the $d$ polarization vectors
\be \label{polarization}
\ve_a^\mu(x)\equiv  \p_a \hat{p}^\mu(x) = \big( x_a , \delta_a^b, - x_a \big) \;.
\ee
These are orthogonal to both $n$ and ${\hat p}$ and satisfy
\begin{equation}
\begin{split}
\ve_a(x) \cdot \ve_b (x) = \delta_{ab}\; , \hspace{.5 in} \ve^a_\mu (x) \ve^\nu_a (x) = \Pi^\nu_\mu (x) \equiv \delta^\nu_\mu + n_\mu {\hat p}^\nu(x) + n^\nu {\hat p}_\mu(x)  \; .
\end{split}
\end{equation}
We also note the property
\begin{equation}
\begin{split}
{\hat p} (x) \cdot \ve_a (x') &=  x_a - x'_a  \; . 
\end{split}
\end{equation}
 The polarization tensors for higher spin representations of the little group can be constructed from the spin-1 polarization forms. For instance, the graviton's polarization tensor is given by
\be
\ve_{\mu \nu}^{ab}(x)=\frac12 \left[	 \ve_\mu^a(x)\ve_\nu^b(x) + \ve_\mu^b (x)\ve_\nu^a(x)\right] - \frac{1}{d} \delta^{ab} \Pi_{\mu\nu}(x)  \; .
\ee

The Fock space of massless scattering states is generated by the algebra of single particle creation and annihilation operators satisfying the standard commutation relations
\be
\big[ \CO_a(p),\CO_b (p')^\dagger \big] = (2\pi)^{d+1} \delta_{a b } (2p^0) \delta^{(d+1)}(\vec{p} - \vec{p}\,' ) \; .
\ee
We can rewrite this relation in terms of our parametrization \eqref{momentum} of the momentum light cone
\be
\big[ \CO_a(\omega,x), \CO_b (\omega', x')^\dagger \big] = 2 (2\pi)^{d+1} \delta_{ab}  \omega^{1-d} \delta( \omega - \omega' ) \delta^{(d)}(x-x')  \; . 
\ee
Note that in terms of the commutation relations of local operators defined on the light cone, the energy direction actually appears space-like rather than time-like.\footnote{This is also the case for the null direction on $\CI$ in asymptotic quantization.} 
The creation and annihilation operators can be viewed as operators inserted at specific points of $\CM_d$ carrying an additional quantum number $\omega$, so that the $\CS$-matrix takes the form of a conformal correlator of primary operators.
In other words, the amplitude with $m$ incoming and $n-m$ outgoing particles\footnote{Here, we have suppressed all other quantum numbers that label the one-particle states, such as the polarization vectors, flavor indices, or charge quantum numbers.}
\be
\CA_{n,m}=\langle p_{m+1}, \dots , p_{n}  | p_1, \dots , p_m\rangle \;  
\ee
can be equivalently represented as a correlation function on $\CM_d$
\be
\CA_{n}= \avg{ \CO_1(p_1) \dots  \CO_{n}(p_n ) }_{\CM_d}  = \avg{ \CO_1(\omega_1, x_1) \dots  \CO_{n}(\omega_n,x_{n}) }_{\CM_d}  \; .
\ee 
In this representation, outgoing states have $\omega >0$ and ingoing states have $\omega <0$. In the rest of this paper, we will freely interchange the notation $p_i \Leftrightarrow (\omega_i,x_i)$ to describe the insertion of local operators.

In appendix \ref{app:coordinates}, we demonstrate that the space-like cut $\CM_d$ of the momentum cone is naturally identified with the cross sectional cuts of $\mathcal{I}^+$.   
We construct the bulk coordinates whose limiting metric on future null infinity is that of a null cone with cross-sectional metric \eqref{Mdmetric}. This provides a holographic interpretation of our construction, recasting scattering amplitudes in asymptotically flat spacetimes in terms of a ``boundary'' theory that lives on $\mathcal{I}^+$.

\section{Lorentz Transformations and the Conformal Group  }\label{sec:lorentztransform}

In this section, we make explicit the map from $(d+2)$-dimensional momentum space $\CS$-matrix elements to conformal correlators on the  Euclidean manifold $\CM_d$. The setup is mathematically similar to the embedding space formalism, although in this case the $(d+2)$-dimensional ``embedding space'' is the physical momentum space rather than merely an auxiliary ambient construction. The Lorentz group $SO(d+1,1)$ with generators $M_{\mu \nu}$ acts linearly on momentum space vectors $p^\mu\in \mathbb{R}^{d+1,1}$. The isomorphism with the conformal group of $\CM_d$ is given by the identifications
\be\label{map}
J_{ab}=M_{ab}\; , \qquad T_a = M_{0,a} - M_{d+1,a}\; , \qquad D = M_{d+1,0}\; , \qquad K_a = M_{0,a} + M_{d+1,a}\; .
\ee
The $J_{ab}$ generate $SO(d)$ rotations, $D$ is the dilation operator, and $T_a$ and  $K_a$ are the generators of translations and special conformal transformations, respectively. These operators satisfy the familiar conformal algebra
\begin{equation}
\begin{split}
[J_{ab}, J_{cd}] &= i(\delta_{ac}J_{bd} + \delta_{bd}J_{ac} - \delta_{bc}J_{ad} - \delta_{ad}J_{bc}) \; , \\
[J_{ab},T_c]  &= i(\delta_{ac}T_b - \delta_{bc}T_a) \; ,\\
[J_{ab}, K_c] &= i(\delta_{ac}K_b - \delta_{bc}K_a)\; ,\\
[D,T_a]  &= -iT_a \; ,\\
[D,K_a]  &= i K_a \; ,\\
[T_a,K_b]  &=-2i(\delta_{ab}D+J_{ab})\; .
\end{split}
\end{equation}
Equation \eqref{map} describes the precise map between the linear action of Lorentz transformations on $p^\mu$ and nonlinear conformal transformations on $(\omega,x)$. The latter amount to transformations $x \to x'(x)$ for which
\be
\frac{\p x'^{c}}{\p x^a}\frac{\p x'^{d}}{\p x^b}\delta_{cd}= \gamma(x)^2 \delta_{ab} 
\ee
along with
\be
\omega \to \omega'=  \frac{\omega}{\gamma(x)} \;.
\ee

Using \eqref{map}, the conformal properties of the operators $\CO(\omega,x)$ (we suppress spin labels for convenience) follow from their Lorentz transformations,
 \begin{align}
 [\CO (p), M_{\mu \nu}]&= \CL_{\mu \nu}\CO (p) + \CS_{\mu \nu} \cdot  \CO(p) \; , \;\;\;\;\;  [\CO (p), P_\mu]=-p_\mu \CO (p) \; ,
 \end{align}
where
\be
\CL_{\mu \nu}= - i \left( p_\mu \frac{\p}{\p p^\nu} - p_\nu \frac{\p}{\p p^\mu} \right) 
\ee
and $\CS_{\mu \nu}$ denotes the spin-$s$ representation of the Lorentz group. We would like to rewrite these relations in a way that manifests the action on $\CM_d$. First, we note that
\be
\frac{\p \omega}{\p p^a}=\frac{2x_a}{1+x^2} \; , \;\;\;\; \omega \frac{\p x^a}{\p p^b}=\delta_b^a-\frac{2x^ax_b}{1+x^2} \; , \;\;\;\; \frac{\p \omega}{\p p^{d+1}}=\frac{2}{1+x^2} \; , \;\;\;\; \omega \frac{\p x^a}{\p p^{d+1}}=-\frac{2x^a}{1+x^2}\; .
\ee
It follows that
\begin{equation}
\begin{split}\label{Lmnexp}
&\CL_{0,a}=ix_a [\omega \p_{\omega} -x^b\p_b] + \frac{i}{2}(1+x^2)\p_a \; , \hspace{.75 in}\CL_{0,d+1}= i[\omega \p_{\omega} -x^a\p_a] \; ,\\
&\CL_{a,d+1}=-ix_a[\omega\p_{\omega}-x^b\p_b]+\frac{i}{2}(1-x^2)\p_a \; , \hspace{.5 in}\CL_{ab}=-i[x_a\p_b -x_b\p_a] \; .
\end{split}
\end{equation}
The action of the spin matrix $\CS_{\mu \nu}$ can be conveniently expressed in terms of the polarization vectors (\ref{polarization}). For instance, the action on a spin-1 state is given by
\be
[\CS_{\mu \nu}]_a{}^{b}=-i \big[ \varepsilon_{\mu a}\varepsilon_{\nu}^{b} -\varepsilon_{\nu a}\varepsilon_{\mu}^{b} \big] + \varepsilon^\rho_a \mathcal{L}_{\mu \nu}\varepsilon_\rho^{b} \; .
\ee
In general one finds
\be\label{Smnexp}
\CS_{0,d+1}=0\;, \hspace{.5 in}\CS_{0a}=\CS_{d+1,a}=x^b\CS_{ab} \;,
\ee
where $\CS_{ab}$ is the representation of the massless little group $SO(d)$. The action of the conformal group on the creation and annihilation operators is then given by
\begin{equation}\label{Commutators}
\begin{split}
&[\CO(\omega,x), T_a]= i\p_a \CO(\omega,x) \; ,\\
 &[\CO(\omega,x), J_{ab}] = -i ( x_a\p_b -x_b\p_a )  \CO(\omega,x) + \CS_{ab} \cdot \CO(\omega,x) \; ,\\
 &[\CO(\omega,x), D]=i ( x^a\p_a -\omega \p_{\omega} ) \CO(\omega,x) \; ,\\
  &[\CO(\omega,x), K_a] = i ( x^2\p_a -  2 x_a x^b \p_b + 2x_a\omega \p_{\omega} ) \CO(\omega,x) + 2x^b \CS_{ab} \cdot \CO(\omega,x) \; .
\end{split}
\end{equation}
We recognize these commutation relations as the defining properties of a spin-$s$ conformal primary operator, with a non-standard dilation eigenvalue
\be
\Delta = -\omega \p_{\omega} \; .
\ee
The fact that $\Delta$ is realized as a derivative simply reflects the fact that the energy eigenstates do not diagonalize the dilation operator, which simply translates the space-like cut $\CM_d$ of the momentum cone along its null direction.\footnote{It is possible to obtain standard conformal primary operators via a Mellin transform, $\CO(\Delta,x) = \int_{\SC} d\omega \omega^{\Delta-1} \CO(\omega,x)$ for some contour $\SC$ in the complex $\omega$ plane.}

\section{Conserved Currents and Soft Theorems}\label{sec:currents}
The operator content and correlation functions of the theory living on $\CM_d$ are highly dependent on the spectrum and interactions of the $(d+2)$-dimensional theory under consideration. However, the universal properties of the $(d+2)$-dimensional $\CS$-matrix are expected to translate into general, model independent features of the ``boundary theory''. In this section, we explore the consequences of the universal soft factorization properties of $\CS$-matrix elements. Soft factorization formulas closely resemble Ward identities, and indeed many soft theorems are known to be intimately related to symmetries of the $\mathcal{S}$-matrix. The existence of the soft theorems should therefore enable one to construct associated conserved currents for the theory living on $\mathcal{M}_d$. The appropriate currents were constructed for $d=2$ in \cite{Strominger:2013lka,He:2015zea,Kapec:2016jld,He:2017fsb,Nande:2017dba,Cheung:2016iub}. Here, we generalize these results to $d>2$.

\subsection{Leading Soft-Photon Theorem and the Conserved $U(1)$ Current }\label{softphoton}

The leading soft-photon theorem is a universal statement about the behavior of $\CS$-matrix elements in the limit that an external photon's momentum tends to zero. It is model independent, exists in any dimension, and states that
\be\label{lsp}
 \avg{ \CO_a(q) \CO_1(p_1)  \dots  \CO_n(p_n) }  ~ \stackrel{q^0 \to 0}{ \longrightarrow}  ~ \sum_{k=1}^n Q_k \frac{\ve_a \cdot p_k}{q \cdot p_k}  \avg{ \CO_1(p_1)  \dots  \CO_n(p_n)  }  \; ,
\ee
where $\CO_a(\omega,x)$ creates an outgoing photon of momentum $p(\omega,x)$ and polarization $\ve^\mu_a(x)$, and $Q_k$ is the charge of the $k$-th particle. We will first define the ``leading soft-photon operator''
\be
S_a(x) = \lim_{\omega \to 0}  \omega \CO_a(\omega,x) \;. 
\ee
This is a conformal primary operator with $(\Delta,s)=(1,1)$. Insertions of this operator are controlled by the leading soft-photon theorem \eqref{lsp} and take the form
\begin{equation}
\begin{split}\label{Sins}
\avg{ S_a(x) \CO_1(\omega_1,x_1) \dots \CO_n(\omega_n,x_n) }
&= \p_a \sum_{k=1}^n Q_k \log \big[(x-x_k)^2\big]   \avg{ \CO_1(\omega_1,x_1) \dots \CO_n(\omega_n, x_n)  }  .
\end{split}
\end{equation}
Note that $S_a(x)$ satisfies
\begin{equation}
\begin{split}\label{Saprop}
\p_a S_b - \p_b S_a = 0 \; 
\end{split}
\end{equation}
identically, without contact terms. 
In even dimensions,\footnote{From this point on, we will consider only the even dimensional case in order to avoid discussion of fractional powers of the Laplacian.} the leading soft-photon theorem is known \cite{He:2014cra,Kapec:2014zla,Mohd:2014oja,Campiglia:2015qka,Kapec:2015ena} to be completely equivalent to the invariance of  the $\CS$-matrix under  a group of angle dependent $U(1)$ gauge transformations with non-compact support.  
In $d=2$, this symmetry is generated by the action of a holomorphic boundary current $J_z$ satisfying the appropriate Kac-Moody Ward identities (see \cite{He:2015zea}).  In higher dimensions, one consequently expects to encounter a conformal primary operator $J_a(x)$ with $(\Delta,s) = (d-1,1)$ satisfying the Ward identity
\begin{equation}
\begin{split}\label{Jins}
 \avg{ \p^b J_b(y) \CO_1(\omega_1,x_1)  \dots  \CO_n(\omega_n,x_n) }
&=  \sum_{k=1}^n Q_k  \delta^{(d)} ( y - x_k )   \avg{ \CO_1(\omega_1,x_1)  \dots  \CO_n(\omega_n, x_n)  } \; . 
\end{split}
\end{equation}

Our goal is to construct the conserved current $J_a(x)$ from the soft operator $S_a(x)$. The inverse problem -- constructing $S_a(x)$ from an operator $J_a(x)$ satisfying (\ref{Jins}) -- is easily solved. Multiplying both sides of  \eqref{Jins} by $\int d^d y \p_a \log [ ( x - y )^2 ] $, we find
\begin{equation}
\begin{split}
& \int d^d y \p_a \log [ ( x - y )^2 ]  \avg{ \p^b J_b(y) \CO_1(\omega_1,x_1)  \dots  \CO_n(\omega_n,x_n) }
\\
&\qquad \qquad \qquad \qquad = \p_a \sum_{k=1}^n Q_k  \log [ ( x - x_k )^2 ]  \avg{ \CO_1(\omega_1,x_1)  \dots  \CO_n(\omega_n, x_n)  } \\
&\qquad \qquad \qquad \qquad = \avg{ S_a(x) \CO_1(\omega_1,x_1)  \dots  \CO_n(\omega_n,x_n) } \;.
\end{split}
\end{equation}
We therefore identify\footnote{Note that this integral expression is insensitive to improvement terms of the form $J_a\to J_a + \p^bK_{[ba]}$ which do not affect the Ward identity (\ref{Jins}). }
\begin{equation}
\begin{split}
S_a(x) =  \int d^d y \p_a \log [ ( x - y )^2 ]  \p^b J_b(y) =  2 \int d^d y  \frac{ \CI_{ab} ( x - y ) }{ ( x - y )^2 }  J^b(y) \;, 
\end{split}
\end{equation}
where $\CI_{ab}(x-y)$ is the conformally covariant tensor
\begin{equation}
\begin{split}
\CI_{ab}(x-y) = \delta_{ab} - 2 \frac{ ( x - y )_a ( x - y )_b }{ ( x - y)^2 } \; .
\end{split}
\end{equation}
This nonlocal relationship between the $\Delta=1$ primary $S_a$ and the $\Delta=d-1$ primary $J_a$   is known as a shadow transform. For a spin-$s$ operator of dimension $\Delta$, the shadow operator is given by \cite{SimmonsDuffin:2012uy}
\begin{equation}
\begin{split}
{\widetilde \CO}_{a_1 \dots a_s} (x) = \delta_{a_1 \dots a_s}^{b_1 \dots b_s} \int d^d y \frac{ \CI_{b_1 c_1} ( x - y ) \dots \CI_{b_s c_s}(x-y) }{ [ ( x - y)^2 ]^{d-\Delta} }  \CO^{c_1 \dots c_s} (y) \; .
\end{split}
\end{equation}
Here, $\delta_{a_1 \dots a_s}^{b_1 \dots b_s}$ is the invariant identity tensor in the spin-$s$ representation,
\begin{equation}
\begin{split}
\delta_{a_1 \dots a_s}^{b_1 \dots b_s} = \delta^{\{a_1}_{\{b_1} \delta^{a_2}_{b_2} \dots \delta^{a_s\}}_{b_s\} }  \; , 
\end{split}
\end{equation}
where the notation $\{ \, ,\}$ denotes the symmetric traceless projection on the indicated indices. 
The shadow transform is the unique integral transform that maps conformal primary operators with $(\Delta,s)$ onto conformal primary operators with $ ( d - \Delta , s )$. Given that $S_a$ has $(\Delta,s)=(1,1)$ while $J_a$ has $(\Delta,s)=(d-1,1)$ it seems natural to expect the appearance of the shadow transform.

The shadow transform is, up to normalization \cite{SimmonsDuffin:2012uy, Rejon-Barrera:2015bpa}, its own inverse\footnote{The spatial integrals involved here are formally divergent and are regulated by the $i\e$-prescription.}
\begin{equation}
\begin{split}\label{shadowinverse}
{\wt {\wt \CO}}_{a_1 \dots a_s} (x) = c(\Delta,s) \CO_{a_1 \dots a_s}(x) \;,  \qquad c ( \Delta,s) = \frac{ \pi^d (\Delta-1)(d-\Delta-1) \Gamma \left( \frac{d}{2} - \Delta \right) \Gamma \left( \Delta - \frac{d}{2} \right) }{  ( \Delta  - 1 + s ) ( d - \Delta  - 1 + s ) \Gamma ( \Delta   )\Gamma ( d - \Delta  ) }  \; . 
\end{split}
\end{equation}
Using this, we can immediately write
\begin{equation}
\begin{split}
S_a(x) =   2 {\wt J}_a(x) \; , \qquad J_a(x) = \frac{1}{ 2 c ( 1 , 1 )   } {\wt S}_a(x) \; . 
\end{split}
\end{equation}
Interestingly, the property \eqref{Saprop} allows one to obtain a local relation between $J_a(x)$ and $S_a(x)$ 
\begin{equation}
\begin{split}
   J_a (x)  &= \frac{1}{   ( 4 \pi )^{d/2}   \Gamma  ( d/2  )  } (-\Box)^{\frac{d}{2}-1} S_a(x) \; . 
\end{split}
\end{equation}
It is straightforward to verify that insertions of $J_a(x)$ are given by
\be
\langle J_a(x) \CO_1(\omega_1,x_1) \dots \CO_n(\omega_n,x_n)\rangle =\frac{\Gamma  ( d / 2 )}{2\pi^{d/2}} \sum_{k=1}^n Q_k\frac{(x-x_k)_a}{ | x - x_k |^d  } \avg{   \CO_1(\omega_1,x_1)\dots \CO_n(\omega_n,x_n) }  \; 
\ee
and satisfy (\ref{Jins}). 

In summary, we find that the leading soft-photon theorem in any dimension implies the existence of a conserved current $J_a(x)$ on the spatial cut $\CM_d$. This current is constructed as the shadow transform of the soft-photon operator $S_a(x)$. This correspondence is reminiscent of a similar construction in AdS/CFT, where again the presence of a massless bulk gauge field produces a dual conserved boundary current. There have been attempts to make this analogy more precise using a so-called ``holographic reduction'' of Minkowski space\cite{deBoer:2003vf,Cheung:2016iub}. The $(d+2)$ dimensional Minkowski space can be foliated by hyperboloids (and a null cone), each of which is invariant under the action of the Lorentz group. Inside the light cone, this amounts to a foliation using a family of Euclidean $\ads_{d+1}$, all sharing an asymptotic boundary given by the celestial sphere.\footnote{This construction seems more natural in momentum space, where it is only the interior of the light cone which is physically relevant, and one never needs to discuss the time-like de Sitter hyperboloids lying outside the light cone.} 
Performing a Kaluza-Klein reduction on the (time-like, non-compact) direction transverse to the $\ads_{d+1}$ slices decomposes the gauge field $A(X)$ in Minkowski space  into a continuum of $\ads_{d+1}$ gauge fields $A_\omega (x)$ with masses $\sim \omega$ ($\omega$ is the so-called \emph{Milne} energy). The $\omega \to 0$ gauge field -- equivalent to the soft limit --  is massless in $\ads_{d+1}$ and therefore induces a conserved current on the $d$-dimensional boundary. The holographic dictionary suggests that in the boundary theory, one has a coupling of the form
\begin{equation}
\begin{split}
\int d^d x S^a (x) J_a(x) \; . 
\end{split}
\end{equation}

The discussion here suggests that a deeper holographic connection (beyond the simple existence of conserved currents) may exist between the theory on $\CM_d$ and dynamics in Minkowski space. While intriguing, much remains to be done in order to elucidate this relationship.  The hypothetical boundary theory is expected to have many peculiar properties, one of which we discuss in section \ref{sec:leadingsoftgraviton}.

\subsection{Subleading Soft-Graviton Theorem and the Stress Tensor}

In the previous section, we demonstrated that the presence of gauge fields in Minkowski space controls the global symmetry structure of the  putative theory on $\mathcal{M}_d$.  As in AdS/CFT, more interesting features arise when we couple the bulk theory to gravity and consider gravitational perturbations. Flat space graviton  scattering amplitudes also display universal behavior in the infrared that is model independent and holds in any dimension. Of particular interest here is the subleading soft-graviton theorem, which states\footnote{We work in units such that $\sqrt{8\pi G}=1$.} 
\be\label{subleadingsoftgravitontheorem}
\lim_{\omega \to 0}(1+\omega \p_\omega)\langle \CO_{a b }(q) \CO_1(p_1) \dots \CO_n(p_n)\rangle =- i\sum_{k=1}^n \frac{\ve^{\mu \nu}_{a b }p_{k\,\mu} q_\rho}{p_k\cdot q}\CJ_k^{\rho \nu}\langle  \CO_1(p_1) \dots \CO_n(p_n)\rangle \;. 
\ee
Here $\CO_{a b }(q)$ creates a graviton with momentum $q$ and polarization $\ve_{\mu \nu}^{a b }(q)$, and  $\CJ_{k \rho \nu}$ is the total angular momentum operator for the $k$-th particle. The operator $(1+\omega \p_\omega)$ projects out the Weinberg pole \cite{Weinberg:1965nx}, yielding a finite $\omega \to 0$ limit.

Returning to the analogy with $\ads_{d+1}/\text{CFT}_d$, one might expect that the bulk soft-graviton is associated to a boundary stress tensor, just as the bulk soft-photon is related to a boundary $U(1)$ current. In a quantum field theory, the stress tensor generates the action of spacetime (conformal) isometries on local operators. As we saw in (\ref{Commutators}), the angular momentum operator $\CJ_{\rho \nu}$ generates these transformations on the local operators on $\mathcal{M}_d$. 
  Therefore, it is natural to suspect that the bulk subleading soft-graviton operator
\be
B_{a b }(x)=\lim_{\omega \to 0}(1+\omega \p_\omega) \CO_{a b }(\omega,x) \;  
\ee
is related to the boundary stress tensor. Such a relationship was derived in four dimensions ($d=2$) in \cite{He:2017fsb,Kapec:2016jld,Cheung:2016iub}. In this section, we generalize the construction to $d> 2$.

Insertions of $B_{ab}(x)$ are controlled by the subleading soft-graviton theorem \eqref{subleadingsoftgravitontheorem} and take the form (see \eqref{Lmnexp} and \eqref{Smnexp} for the explicit forms of the orbital and spin angular momentum operators)
\begin{equation}
\begin{split}
\langle B_{ab}(x)\CO_1(\omega_1,x_1)  \dots \CO_n(\omega_n,x_n)\rangle &= \sum_{k=1}^n \left[	\CP^c{}_{ab}(x-x_k)\p_{x_k^c}	+ \frac{1}{d} \p_c \CP^c{}_{ab}(x-x_k)\omega_k \p_{\omega_k} \right. \\
&\left.   -\frac{i}{2}\p^{[c}\CP^{d]}{}_{ab}(x-x_k)\CS_{kcd} 	\right]\langle \CO_1(\omega_1,x_1)\dots \CO_n(\omega_n,x_n)\rangle \; ,
\end{split}
\end{equation}
where
\be
	\CP^c{}_{ab}(x)=\frac12 \left[x_a\delta_b^c +x_b \delta_a^c +\frac{2}{d}x^c \delta_{ab}-\frac{4}{x^2}x^cx_ax_b\right] \; .
\ee
One can check that
\begin{equation}
\begin{split}
\p_{\{c}\CP_{d\}ab}(x)=   \CI_{\{\underline{a}\{c} (x) \CI_{d\}\underline{b}\}} (x) \; .  
\end{split}
\end{equation}

As in section \ref{softphoton},  
it is easiest to first determine $B_{ab}$ in terms of $T_{ab}$. 
Recall that the Ward identities for the energy momentum tensor of a CFT$_d$ take the form \cite{DiFrancesco:1997nk}
\begin{align} \label{div}
\avg{ \p^dT_{dc}(y) \CO_1(\omega_1,x_1) \dots \CO_n(\omega_n,x_n) } &= -\sum_{k=1}^n \delta^{(d)}(y-x_k)\p_{x_k^c}\avg{  \CO_1(\omega_1,x_1) \dots \CO_n(\omega_n,x_n) } \; ,\\
\label{trace}
\avg{ T^c{}_{c}(y) \CO_1(\omega_1,x_1) \dots \CO_n(\omega_n,x_n) } &= \sum_{k=1}^n \delta^{(d)}(y-x_k)\omega_k\p_{\omega_k}\avg{ \CO_1(\omega_1,x_1) \dots \CO_n(\omega_n,x_n)} \; ,
\\
\label{antisym}
\avg{ T^{[cd]}(y) \CO_1(\omega_1,x_1) \dots \CO_n(\omega_n,x_n) } &= -\frac{i}{2}\sum_{k=1}^n \delta^{(d)}(y-x_k)\CS_k^{cd}\avg{  \CO_1(\omega_1,x_1) \dots \CO_n(\omega_n,x_n) } \; .
\end{align}
Multiplying (\ref{div}) by $- \int d^dy \CP^c{}_{ab}(x-y)$, (\ref{trace}) by $\frac{1}{d} \int d^dy \p_c \CP^c{}_{ab}(x-y)$, (\ref{antisym}) by \\$\int d^dy \p^c \CP^d{}_{ab}(x-y)$, and taking the sum, one finds
\be
\begin{split}\label{Babrel}
B_{ab}(x) &= -  \int d^d y  \p_{\{c} \CP_{d\}ab} (x - y)   T^{cd}(y)  \\
&=  -\int d^dy \CI_{\{\underline{a}\{c} (x-y) \CI_{d\}\underline{b}\}} (x-y)T^{cd}(y)  \\
&= - {\wt T}_{\{ ab \}}(x) \;.
\end{split}
\ee
Once again, the soft operator appears as the shadow transform of a conserved current. The relationship could have been guessed from the outset based on the dimensions of $B_{ab}$ and $T_{\{ab\}}$.\footnote{Note that only the symmetric traceless  part of the stress tensor appears in this dictionary since the graviton lies in the symmetric traceless representation of the little group. The trace term may be related to soft-dilaton theorems. }

Having derived \eqref{Babrel}, we can now invert the shadow transform to find
\begin{equation}
\begin{split}\label{Tabrel}
T_{\{ ab \}}(x) =- \frac{1}{ c ( 0 , 2 ) } {\wt B}_{ab}(x) \; .
\end{split}
\end{equation}

The shadow relationship between the soft operator $B_{ab}$ and the energy momentum tensor is again suggestive of a coupling
\be
\int d^dx B^{ab}(x)T_{ab}(x)
\ee
in some hypothetical dual formulation of asymptotically flat gravity: the  soft-graviton creates an infinitesimal change in the boundary metric, sourcing the operator $T_{ab}$. In \cite{Kapec:2016jld}, it was viewed as a puzzle that the energy momentum tensor appears non-local when written in terms of the soft-modes of the four dimensional gravitational field. Here we see that this is essentially the consequence of a linear response calculation, and that the non-locality is actually the only one allowed by conformal symmetry.

As in \cite{Kapec:2016jld}, it is possible to derive a local differential equation for $T_{\{ab\}}$ in even dimensions. We first define the following derivative operator
\begin{equation}
\begin{split}
\mdd^a O_{ab} &\equiv \frac{1}{2(4\pi)^{d/2}\Gamma(d/2+1)} \big[(-\Box)^{d/2}\p^a  O_{ab}  + \frac{d}{d-1}\p_b (-\Box)^{d/2-1}\p^e\p^f  O_{ef}  \big] \; . 
\end{split}
\end{equation}
One can check that 
\be
\mdd^a \CP^c{}_{ab}  = -\delta_b^c \delta^{(d)}(x)  \; . 
\ee
Then, acting on the first equation of \eqref{Babrel} with $\mdd^a$, we find
\be
\begin{split}
\mdd^a B_{ab}(x) &=  \p^a T_{\{ab\}}(y) \; .
\end{split}
\ee

\subsection{Leading Soft-Graviton Theorem and Momentum Conservation}\label{sec:leadingsoftgraviton}

In the previous two subsections, we have avoided the discussion of currents related to the leading soft-graviton theorem. This soft theorem is associated to spacetime translational invariance (and more generally to BMS supertranslations \cite{Strominger:2013jfa,Campiglia:2015kxa,He:2014laa,Kapec:2015vwa}). For scattering amplitudes in the usual plane wave basis, this symmetry is naturally enforced by a momentum conserving Dirac delta function. In our discussion above, we have chosen to make manifest the Lorentz transformation properties of the scattering amplitude. Consequently, translation invariance, or global momentum conservation, is unwieldy in our formalism. In fact, it must somehow appear as a non-local constraint on the correlation functions on $\CM_d$, since arbitrary operator insertions corresponding to arbitrary configurations of incoming and outgoing momenta will in general violate momentum conservation. The difficulty can  also be seen at the level of the symmetry algebra. Momentum conservation cannot arise simply as a global $\mathbb{R}^{d+1,1}$ symmetry of the CFT$_d$, since the associated conserved charges do not commute with the conformal (Lorentz) group. In light of this it is not clear that our construction can really be viewed as a local conformal field theory living on $\CM_d$.\footnote{However, it is also not clear that we should expect a local QFT dual to asymptotically flat quantum gravity. The flat space Bekenstein-Hawking entropy is always super-Hagedorn in $d\geq4$. The high energy density of states grows faster than in any local theory.} We have tried, unsuccessfully, to find a natural set of operators whose shadow reproduces the leading  soft-graviton theorem\footnote{It has also been suggested \cite{Shu-Heng} that translational invariance of the $\mathcal{S}$-matrix is realized through null state relations of boundary correlators  rather than through local current operators, since the former are typically non-local constraints on CFT correlation functions.}
\be
\lim_{\omega \to 0}\omega \langle \CO_{a b }(q) \CO_1(p_1)\dots  \CO_n(p_n) \rangle = \omega \sum_{k=1}^n  \frac{\ve_{\mu \nu}^{a b }p_{k}^\mu p_k^\nu}{p_k\cdot q} \langle  \CO_1(p_1)\dots  \CO_n(p_n) \rangle \; .
\ee
The soft operator
\be
G_{ab}(x)=\lim_{\omega \to 0}\omega \CO_{a b }(\omega,x)
\ee
has insertions given by
\be
\langle G_{ab}(x)\CO_1(\omega_1,x_1)\dots  \CO_n(\omega_n,x_n) \rangle = \frac{2}{d}\sum_{k=1}^n \omega_k \CI_{ab}^{(d)}(x - x_k)\langle \CO_1(\omega_1,x_1)\dots  \CO_n(\omega_n,x_n) \rangle \; ,
\ee
where 
\be
\CI_{ab}^{(d)} (x ) = \delta_{ab} - d \frac{x_ax_b}{x^2} \; .
\ee
 One finds that the operator
\be
U_a = \frac{ 1 }{  ( 4 \pi ) ^{d/2}   \Gamma \left( d/ 2 \right)  ( d - 1 )   } (-\Box)^{\frac{d}{2}-1}\p^bG_{ba}
\ee
satisfies
\be\label{Uains}
\langle \p^a U_{a}(x) \CO_1(\omega_1,x_1)\dots  \CO_n(\omega_n,x_n) \rangle = - \sum_{k=1}^n \omega_k \delta^{(d)} ( x - x_k )  \langle \CO_1(\omega_1,x_1)\dots  \CO_n(\omega_n,x_n) \rangle \; .
\ee 
Thus, $U_a(x)$ satisfies the current Ward identity corresponding to ``energy'' conservation. However, since $\omega$ is not a scalar charge, the current $U_a(x)$ is not a primary operator (though it has a well-defined scaling dimension $\Delta = d$). 
Acting on \eqref{Uains} with $ - \frac{2}{d} \int d^d x  \CI_{ab}^{(d)}( y - x)$, one finds
\begin{equation}
\begin{split}\label{Gabrel}
G_{ab}(x) &= - \frac{2}{d} \int d^dy \, \CI_{ab}^{(d)}(x-y)\p^cU_c(y) \;. \\
\end{split}
\end{equation}
Unlike the $U(1)$ current and the stress tensor, the leading soft-graviton ``current'' is not related to the soft operator $G_{ab}$ through a shadow transform. It may be possible to interpret \eqref{Gabrel} as some other (conformally) natural non-local transform of $U_c(x)$, but we do not pursue this here.\footnote{The shadow transform $(\Delta,s) \to (d-\Delta,s)$ is related to the  $\mzz_2$ symmetry of the quadratic and quartic Casimirs of the conformal group $c_2 =   \Delta(d-\Delta) + s(2-d-s) $, $c_4 = -s(2-d-s)(\Delta-1)(d-\Delta-1)$. $c_2$ and $c_4$ are also invariant under another $\mzz_2$ symmetry under which $(\Delta,s)\to (1-s,1-\Delta)$. Equation \eqref{Gabrel} may be the integral representation of a shadow transform followed by the second $\mzz_2$ transform which maps $(d+1,0) \to (-1,0) \to (1,2)$.}

\section{Conclusion}\label{sec:comments}
In this paper we have taken steps to recast the $(d+2)$-dimensional $\CS$-matrix as a collection of celestial correlators, but many open questions remain. Our analysis relied on symmetry together with the universal behavior of the $\mathcal{S}$-matrix in certain kinematic regimes. It would be interesting to analyze the consequences of other universal properties of the $\mathcal{S}$-matrix for the celestial correlators. The analytic structure and unitarity of the $\mathcal{S}$-matrix should be encoded in properties of these correlation functions, although the mechanism may be subtle. It seems likely that the collinear factorization of the $\mathcal{S}$-matrix could be used to define some variant of the operator product expansion for local operators on the light cone. Although this paper only addressed single soft insertions, double soft limits, appropriately defined, could be used to define OPE's between the conserved currents and stress tensors. We expect supergravity soft theorems to yield a variety of interesting operators, including a supercurrent. Finally, the interplay of momentum conservation with the CFT$_d$ structure requires further clarification. We leave these questions to future work.

\section*{Acknowledgements}
We are  grateful to  Sabrina Pasterski, Abhishek Pathak, Ana-Maria Raclariu, Shu-Heng Shao, Andrew Strominger, Xi Yin, and Sasha Zhiboedov for discussions. This work was supported in part by DOE grant DE-SC0007870 and the Fundamental Laws Initiative at Harvard. PM gratefully acknowledges support from DOE grant DE-SC0009988.

\appendix
\section{Spacetime Picture\label{app:coordinates}}
In this appendix we construct coordinates for Mink$_{d+2}$ whose limiting metric on $\mathcal{I}^+$ has cross sectional cuts given by $\mathcal{M}_d$. Consider the coordinate transformation from the flat Cartesian coordinates $X^\mu$ to the coordinates $(u,r,x^a)$ given by
\be
X^\mu(u,r,x^a)= rp^\mu(x^a) + uk^\mu(x^a) \; ,
\ee
where $p^\mu(x^a)$ is given by (\ref{momentum}) with $\omega$ set to one. We have
\be
dp^\mu dp_\mu = \Omega^2(x)  dx^a dx_a \; .
\ee
If the vector $k^\mu(x^a)$ is chosen to satisfy  
\be
\p_ak^\mu=0 \; , \hspace{.5 in} k^\mu \p_a p_\mu = 0 \; ,
\ee
then one finds
\be
dX^\mu dX_\mu = k^2 du^2 + 2(p\cdot k) dudr + r^2 dp^\mu dp_\mu \; .
\ee
Since neither $p^\mu$ nor $k^\mu$ scales with $r$, the limit  $r\to \infty$ with $u$ fixed yields a degenerate metric on $\mathcal{I}^+$ with cross-sectional metric 
\be
ds_{}^2 =  r^2\Omega^2(x) \delta_{ab}dx^adx^b = r^2ds^2_{\mathcal{M}_d} \; .
\ee
For instance, the flat metric on $\mathcal{M}_d$ corresponds to the choice
\be
p^\mu(x)=  \left(\frac{1+x^2}{2},x^a,\frac{1-x^2}{2} \right) \; , \hspace{.5 in} k^\mu = \frac12 \left[1,0^a,-1 \right] \; ,
\ee
which yields the familiar coordinate transformation
\be\label{flatcoord}
X^\mu(u,r,x^a)= \left[	\frac{u+r(1+x^2)}{2},rx^a,\frac{r(1-x^2)-u}{2}		\right] \; 
\ee
and a metric of the form
\be
ds^2=-dudr + r^2 \delta_{ab}dx^adx^b \; .
\ee
In order to achieve cross sectional cuts of $\mathcal{I}^+$ which are metrically $S^d$, one chooses
\be
p^\mu(x)= \frac{2}{1+x^2} \left(\frac{1+x^2}{2},x^a,\frac{1-x^2}{2} \right) \; , \hspace{.5 in}  k^\mu=\left[ 1, 0^a, 0 \right] \; . 
\ee
The coordinate transformation is then
\be
X^\mu(u,r,x^a)=\left[ u+r, \frac{2rx_a}{1+x^2}, r \frac{1-x^2}{1+x^2}		\right] 
\ee 
and the spacetime metric is 
\be
ds^2=-du^2-2dudr + \frac{4r^2}{(1+x^2)^2}\delta_{ab}dx^adx^b \; .
\ee
In order to make the relationship between $\mathcal{M}_d$ and $\mathcal{I}^+$ even more explicit, consider the flat coordinate system \eqref{flatcoord}. A massless field in the plane wave basis takes the form
\be
\Phi_{\mu_1 \dots \mu_s}(X) = \sum_{a_i}\int \frac{d^{d+1}q}{(2\pi)^{d+1}}\frac{1}{2\omega_q}\ve_{\mu_1 \ldots \mu_s}^{a_1\ldots a_s}(q) \left[	\CO_{a_1\dots a_s}(q)e^{iq \cdot X} + \CO^\dagger_{a_1\dots a_s}(q) e^{-iq\cdot X} \right] \; .
\ee
To perform an asymptotic analysis near $\mathcal{I}^+$, one considers the limit $r \to \infty$ with $u$ fixed, so that  $X\to r p^\mu(x^a)$. In this limit, the argument of the exponential 
\be
ir\omega_q q(y_a)\cdot p (x_a)=-\frac{i}{2}r\omega_q(x-y)^2
\ee
is large so that the exponential is rapidly oscillating. At leading order in $\frac{1}{r}$, the only momenta that contribute to the integral are those for which the phase is stationary, i.e. for which $x=y$. Therefore in the large-$r$ light-like limit, one effectively trades the transverse coordinates on $\mathcal{I}^+$ for the momentum coordinates on $\CM_d$.

Armed with this knowledge we can further elaborate on the results of section \ref{sec:currents}. 
The soft-photon operator $S_a(x)$ is related to the boundary current $J_a(x)$ through a differential equation of the form
\be \label{source}
  (-\Box)^{\frac{d}{2}-1} S_a(x) =  ( 4 \pi )^{d/2}   \Gamma  ( d/2  )J_a(x) \; .
\ee 
The physical picture is clear: charged particles passing through $\mathcal{I}^+$ act as a source $J_{a}(x)$ for the soft radiation $S_a(x)$ (see \cite{Kapec:2014zla} for  relevant expressions relating \eqref{source} to the soft charge for large $U(1)$ gauge transformations). 
Similar statements apply to the gravitational case \cite{Kapec:2015vwa}. Energetic particles passing through $\mathcal{I}^+$ act as an effective source (the boundary energy momentum tensor $T_{ab}$) for  soft-graviton radiation $B_{ab}(x)$.

\bibliography{km-references}
\bibliographystyle{utphys}

\end{document}